\documentstyle[12pt]{article}
\begin{document}

\hfill DUKE-CGTP-2001-02

\hfill hep-th/0102197

\vspace*{1.5in}

\begin{center}

{\large\bf Stacks and D-Brane Bundles }

\vspace{1in}

Eric Sharpe \\
Department of Physics \\
Box 90305 \\
Duke University \\
Durham, NC  27708 \\
{\tt ersharpe@cgtp.duke.edu} \\

$\,$ \\

\end{center}

In this paper we describe explicitly how the twisted ``bundles''
on a D-brane worldvolume in the presence of a nontrivial B field,
can be understood in terms of sheaves on stacks.  We also take
this opportunity to provide the physics community with a readable
introduction to stacks and generalized spaces.

\begin{flushleft}
February 2001 
\end{flushleft}

\newpage

\tableofcontents

\newpage

\section{Introduction}

In the presence of a nontrivial NS $B$ field, the bundle on 
a D-brane worldvolume is not really a bundle at all, but rather
is twisted.  This was pointed out by, for example, \cite{freeded},
and is due to the fact that under a gauge transformation
\begin{displaymath}
B \: \mapsto \: B \: + \: d \Lambda
\end{displaymath}
the Chan-Paton gauge field $A$ necessarily transforms as
\begin{displaymath}
A \: \mapsto \: A \: - \: \Lambda I
\end{displaymath}
(where $I$ is the $N \times N$ identity, for a $U(N)$ gauge theory).

As a result, if we describe the $B$ field in generality in terms
of local 2-forms $B^{\alpha}$ together with overlap data
$A^{\alpha \beta}$, $h_{\alpha \beta \gamma}$ (with respect
to some open cover $\{ U_{\alpha} \}$), related as
\begin{eqnarray*}
B^{\alpha} \: - \: B^{\beta} & = & d A^{\alpha \beta} \\
A^{\alpha \beta} \: + \: A^{\beta \gamma} \: + \: A^{\gamma \alpha}
& = & d \log h_{\alpha \beta \gamma} \\
\delta h_{\alpha \beta \gamma} & = & 1
\end{eqnarray*}
then the transition functions $g_{\alpha \beta}$ of the D-brane
worldvolume ``bundle'' no longer completely close on triple overlaps,
but rather
\begin{displaymath}
g_{\alpha \beta} g_{\beta \gamma} g_{\gamma \alpha} \: = \: 
h_{\alpha \beta \gamma} I
\end{displaymath}

Now, one can certainly ask what sort of formal structure should be used
to describe such ``bundles.''  

One approach, popular in noncommutative geometry circles, is to think
of these as modules over Azumaya algebras \cite{kapustin}
(see also \cite{bm,caldararu}).
However, this description has two drawbacks:
\begin{enumerate}
\item First, Azumaya algebras are only relevant to describing
$B$-field configurations whose curvature $H$ is a torsion element
of $H^3({\bf Z})$.  One would like a more general understanding.
\item Second, the Azumaya algebra description is closely
tied to noncommutative geometry, and one would like a description
that is useful away from the Seiberg-Witten noncommutative geometry
limit.
\end{enumerate}

In this technical note we shall point out that the D-brane worldvolume
``bundles'' can also be understood in terms of sheaves on stacks,
where the stacks in question arise as the formal structures describing
the $B$-field (in the same sense that a bundle is a formal structure describing
a gauge field).

More generally, there is lore in the mathematics community that
stacks and noncommutative geometry are not unrelated.
The work in this paper -- the fact that D-brane ``bundles,''
at least for torsion $H$, can be understood either in terms
of noncommutative geometry (modules over Azumaya algebras) or as 
sheaves on stacks -- is just one aspect of the correspondence.
We hope to report in detail on this correspondence in future work.

Because stacks are not familiar to most physicists, we have included
a short review of relevant definitions and aspects.  Although stacks
have gained a fearsome reputation in some quarters, we hope that our
presentation should help dispell myths that stacks are necessarily
difficult to work with.

We begin in section~\ref{gpoid} with a review of groupoids.
The precise technical definition of ``groupoid'' unfortunately seems
to vary between authors; our use is in the sense of \cite{vistoli,gomez},
and essentially is the same thing as a presheaf of categories.
In section~\ref{presf} we describe presheaves on groupoids.
In section~\ref{site} we describe how one can put a topology on
a groupoid, which is necessary if one wants to talk about sheaves
on groupoids, not just presheaves.  In section~\ref{stk} we describe
stacks, which are special kinds of groupoids.  In essence,
if we view a groupoid as a presheaf of categories, then a stack is
a sheaf of categories.  Finally in section~\ref{sfstk} we describe
sheaves on stacks, and in particular describe specifically how the
twisted ``bundles'' on D-brane worldvolumes can be understood as
sheaves on stacks (where the stacks in question describe the $B$-field).
We conclude in section~\ref{concl}.  We have also included an appendix
describing explicitly how groupoids are related to ``presheaves of
categories,'' as described in, for example, \cite{brylinski,breen,dt2}.

\section{Groupoids}     \label{gpoid}

\subsection{Definition of groupoid}

Let ${\cal U}$ be the category of
open sets on $X$ (i.e., objects are open sets, and morphisms
are inclusions).

A {\it groupoid}\footnote{The definition of ``groupoid'' unfortunately
varies slightly from author to author; we are using the term in the sense
of \cite{vistoli}.} over a space $X$  
\cite{vistoli,gomez} is defined to be a category ${\cal F}$
together with a functor $P_{{\cal F}}: {\cal F} \rightarrow {\cal U}$,
obeying the following two axioms:
\begin{enumerate}
\item If $\rho: U \hookrightarrow V$ is a morphism in ${\cal U}$
and $\eta$ is an object of ${\cal F}$ with $p_{{\cal F}}(\eta) = V$, 
then there exists
an object $\xi \in \mbox{Ob } {\cal F}$ and a morphism $f: \xi \rightarrow \eta$
such that $p_{{\cal F}}(\xi) = U$ and $p_{{\cal F}}(f) = \rho$.
\item If $\phi: \xi \rightarrow \zeta$ and $\psi: \eta \rightarrow \zeta$
are morphisms in ${\cal F}$ and $h: p_{{\cal F}}(\xi) \rightarrow 
p_{{\cal F}}(\eta)$
is a morphism such that $p_{{\cal F}}(\psi) \circ h = p_{{\cal F}}(\phi)$,
then there exists a unique morphism $\chi: \xi \rightarrow \eta$
such that $\psi \circ \chi = \phi$ and $p_{{\cal F}}(\chi) = h$.
\end{enumerate}

Strictly speaking, references \cite{vistoli,gomez} defined
groupoids in algebraic geometry.  Instead of working with a category
${\cal U}$ of open sets on $X$, those references replaced ${\cal U}$
with the category of $X$-schemes.  However, aside from that replacement,
the definitions are identical.  We should also point out that 
essentially the same approach to stacks has previously been used in
\cite{brylinski,breen,dt2}.  More generally, although stacks are
often used by algebraic geometers and so are often written in reference
to schemes, stacks do not have anything to do with schemes {\it per se},
and can be described at a purely topological level.  
See \cite{brylinski,breen,dt2,fantechi,teleman} for discussions of
such topological stacks. 

Intuitively, what is a groupoid?  Although it is not immediately
obvious, a groupoid is equivalent to a ``presheaf of categories,''
i.e., a prestack as defined in \cite{dt2}.
We discuss this equivalence in detail in
appendix~\ref{gpdeqprest}.  We shall have little use for this
point of view, but it does give a more intuitive picture of groupoids.

As a technical aside, note that the pair $(\xi, f)$ determined in
the first groupoid axiom differs from any other pair $(\xi', f')$
satisfying the same
constraint by a unique isomorphism $\chi: \xi \rightarrow \xi'$, 
according to the second axiom.
Since $\xi$ is determined uniquely (up to unique isomorphism), 
$\xi$ is commonly denoted $\rho^* \eta$.  Also note that there
is a canonical morphism $\rho^* \eta \rightarrow \eta$.

As another technical aside, note that if $\alpha$ is a morphism in ${\cal F}$
such that $p_{{\cal F}}(\alpha) = \mbox{Id}$, then
$\alpha$ is necessarily an isomorphism.
Write $\alpha: \xi \rightarrow \zeta$, then apply the second
groupoid axiom to the maps $\psi \equiv \alpha: \xi \rightarrow
\zeta$ and $\phi \equiv \mbox{Id}:  \zeta \rightarrow \zeta$
to find that there exists a unique map $\chi$ such that
$\alpha \circ \chi = \mbox{Id}$.  Using almost the same argument
one can show that $\chi' \circ \alpha = \mbox{Id}$ for some $\chi'$,
and by composition, $\chi = \chi'$.  Hence
$\alpha$ is invertible, and hence is an isomorphism.
Conversely, if $\alpha$ is an isomorphism, we have that $p_{{\cal F}}(\alpha)
= \mbox{Id}$ (since we have defined groupoids over ordinary spaces).


One commonly denotes by ${\cal F}(U)$, $U$ an open set on $X$
and ${\cal F}$ a groupoid on $X$, the subcategory of ${\cal F}$
defined by
\begin{enumerate}
\item Objects are $\eta \in \mbox{Ob } {\cal F}$ such that
$p_{{\cal F}}(\eta) = U$
\item Morphisms are morphisms $f: \xi \rightarrow \eta$
such that $\xi, \eta \in \mbox{Ob } {\cal F}(U)$
and $p_{{\cal F}}(f) = \mbox{Id}_U$.
\end{enumerate}
Note this means that morphisms in ${\cal F}(U)$ are invertible.

Given a groupoid ${\cal F}$, we can define a presheaf of
sets $\underline{\mbox{Hom}}_U(x,y)$ for any open $U \subseteq X$
and any $x, y \in \mbox{Ob } {\cal F}(U)$, as follows:
\begin{enumerate}
\item For any open $V \subseteq U$, the set of sections is
the set $\mbox{Hom}_{{\cal F}(V)}(x|_V, y|_V)$.
\item For any inclusion map $\rho: U_2 \hookrightarrow U_1$
of open sets in $U$, define the restriction map 
\begin{displaymath}
\rho^*: \mbox{Hom}_{{\cal F}(U_1)}\left(x|_{U_1}, y|_{U_1}\right)
\: \longrightarrow \:
\mbox{Hom}_{{\cal F}(U_2)}\left(x|_{U_2}, y|_{U_2}\right)
\end{displaymath}
as follows.
Let $f: x|_{U_1} \rightarrow y|_{U_1}$ be an element of the
set (i.e., a morphism).
Define $\rho^* f$ to be the unique map that makes the following
diagram commute:
%
%
\end{enumerate}
\begin{displaymath}
\begin{array}{ccccc}
x & & = & & x \\
\uparrow & & & & \uparrow \\
x|_{U_1} & \leftarrow & x|_{U_1}|_{U_2} & \stackrel{*}{\rightarrow} &
x|_{U_2} \\
\makebox[0pt][r]{ $\scriptstyle{ f}$ } \downarrow & & \downarrow
\makebox[0pt][l]{ $\scriptstyle{ f|_{U_2} }$ } & & 
\downarrow \makebox[0pt][l]{ $\scriptstyle{ \rho^* f}$ } \\
y|_{U_1} & \leftarrow & y|_{U_1}|_{U_2} & \stackrel{*}{\rightarrow} &
y|_{U_2} \\
\downarrow & & & & \downarrow \\
y & & = & & y 
\end{array}
\end{displaymath}
where unmarked arrows are canonical, $f|_{U_2}$ is the unique morphism
making the middle left square commute (whose existence follows
from the second groupoid axiom), and the arrows marked with a $*$
are the unique morphisms that make the top and bottom squares commute
(whose existence follows from the second groupoid axiom).

Note that the presheaf above is not quite defined uniquely,
since the restrictions $x|_V$ are only defined up
to unique isomorphism.
However, it is straightforward to check that those unique isomorphisms
define an isomorphism of presheaves,
so the presheaf 
$\underline{\mbox{Hom}}_U(x,y)$ is defined up to
unique isomorphism.


\subsection{Examples}

Let $X$ be a topological space, and ${\cal U}$ the category of
open sets on $X$.  Then ${\cal U}$, together with the identity
map ${\cal U} \rightarrow {\cal U}$, forms a trivial example
of a groupoid on $X$.  Sometimes we shall simply use $X$ itself to denote
this groupoid trivially associated to $X$.

Let ${\cal S}$ be a presheaf of sets on a space $X$.
Then ${\cal S}$ defines a groupoid on $X$, as follows:
\begin{enumerate}
\item Objects are pairs $(U, \xi)$, where $U$ is an open subset of $X$,
and $\xi \in {\cal S}(U)$.
\item Morphisms $(U, \xi) \rightarrow (V, \zeta)$ are inclusion maps
$\rho: U \hookrightarrow V$ such that $\xi = \rho^* \zeta$.
\end{enumerate}
It is straightforward to check that this defines a groupoid on $X$. 
(Indeed, presheaves of sets furnish trivial examples of prestacks,
as noted in \cite{dt2}.)

\subsection{Morphisms of groupoids}

A morphism of groupoids ${\cal F} \rightarrow {\cal G}$ over a space $X$
is defined
to be a functor $F: {\cal F} \rightarrow {\cal G}$ such that
$p_{{\cal G}} \circ F = p_{{\cal F}}$, i.e.,
\begin{displaymath}
\begin{array}{ccc}
{\cal F} & \stackrel{ F }{\longrightarrow} & {\cal G} \\
\makebox[0pt][r]{ $\scriptstyle{ p_{{\cal F}} }$ } \downarrow & &
\downarrow \makebox[0pt][l]{ $\scriptstyle{ p_{{\cal G}} }$ } \\
X & = & X
\end{array}
\end{displaymath}
commutes.

We shall see in appendix~\ref{gpdeqprest} that a morphism of groupoids
is equivalent to a Cartesian functor between the associated prestacks.
Also, a natural transformation between two morphisms of groupoids
is equivalent to a 2-arrow between associated Cartesian functors.

\subsection{Yoneda lemma}

Let ${\cal F}$ be a groupoid on a space $X$,
and $U$ an open set on $X$.
Let $\mbox{Hom}(U, {\cal F})$ denote the category of
groupoid morphisms $U \rightarrow {\cal F}$, where we
let $U$ also denote the groupoid canonically associated to the space $U$.

Define a functor $u: \mbox{Hom}(U, {\cal F}) \rightarrow {\cal F}(U)$
as follows:
\begin{enumerate}
\item Objects:  Let $f: U \rightarrow {\cal F}$ be a groupoid
morphism (i.e., an object of $\mbox{Hom}(U, {\cal F})$).
Define
\begin{displaymath}
u: \: f \: \mapsto \: f(U)
\end{displaymath}
\item Morphisms:  Let $\eta: f \Rightarrow g$ be a natural transformation
(i.e., a morphism in $\mbox{Hom}(U, {\cal F})$) between groupoid
morphisms $f, g: U \rightarrow {\cal F}$.
Define
\begin{displaymath}
u: \: \eta \: \mapsto \: \{ \, \eta(U): \: f(U) \: \longrightarrow \:
g(U) \, \}
\end{displaymath}
\end{enumerate}
With these definitions, $u: \mbox{Hom}(U, {\cal F}) \rightarrow
{\cal F}(U)$ is a well-defined functor.

In this section, we shall show that this functor
$u: \mbox{Hom}(U, {\cal F}) \rightarrow {\cal F}(U)$ is an equivalence
of categories.  More precisely, we shall explicitly write
down a functor $v: {\cal F}(U) \rightarrow \mbox{Hom}(U, {\cal F})$
such that $u \circ v \cong \mbox{Id}$ and $v \circ u \cong \mbox{Id}$.

Define a functor $v: {\cal F}(U) \rightarrow \mbox{Hom}(U, {\cal F})$
as follows.
\begin{enumerate}
\item Objects:  Let $\eta \in \mbox{Ob } {\cal F}(U)$.
Define a groupoid morphism $f_{\eta}: U \rightarrow {\cal F}$ as follows:
\begin{enumerate}
\item Objects:  First, $f_{\eta}: U \mapsto \eta$,
and if $\rho: V \hookrightarrow U$ is inclusion of open sets,
then $f_{\eta}: V \mapsto \rho^* \eta$.
\item Morphisms: Let $\rho: V_1 \hookrightarrow V_2$ be inclusion of
open sets.  From the second groupoid axiom, there exists a unique
morphism $\chi_{12}: \eta|_{V_1} \rightarrow \eta|_{V_2}$ such
that $p_{{\cal F}}(\chi_{12}) = \rho$, and
the composition $\eta|_{V_1} \stackrel{ \chi_{12} }{ \longrightarrow }
\eta|_{V_2} \longrightarrow \eta$ equals the natural map
$\eta|_{V_1} \rightarrow \eta$.
Define
\begin{displaymath}
f_{\eta}: \: \rho \: \mapsto \: \chi_{12}
\end{displaymath}
\end{enumerate}
With these definitions, $f_{\eta}: U \rightarrow {\cal F}$ is a well-defined
groupoid morphism.
Thus, we define
\begin{displaymath}
v: \: \eta \: \mapsto \: f_{\eta}
\end{displaymath}
(In passing, note that $f_{\eta}$ is only defined up to isomorphism,
because $\rho^* \eta$ is only unique up to unique isomorphism.
It is straightforward to check that this implies that our
functor $v$ is only defined up to an invertible natural transformation.)
\item Morphisms:  Let $\eta_1, \eta_2 \in \mbox{Ob } {\cal F}(U)$,
and $g: \eta_1 \rightarrow \eta_2$ a morphism in ${\cal F}(U)$.
Define a natural transformation $\tilde{g}: f_{\eta_1} \Rightarrow
f_{\eta_2}$ as follows:
for every open $V \subseteq U$, define
$\tilde{g}(V): f_{\eta_1} \rightarrow f_{\eta_2}$ to be the
unique morphism given by the second groupoid axiom such that the
following diagram commutes:
\begin{displaymath}
\begin{array}{ccc}
f_{\eta_1}(V) = \eta_1|_V & \stackrel{ \tilde{g}(V) }{\longrightarrow} &
f_{\eta_2}(V) = \eta_2|_V \\
\downarrow & & \downarrow \\
\eta_1 & \stackrel{g}{\longrightarrow} & \eta_2
\end{array}
\end{displaymath}
With this definition, it is straightforward to check
that $\tilde{g}$ is a natural transformation.
Then, we define
\begin{displaymath}
v: \: g \: \mapsto \: \tilde{g}
\end{displaymath}
\end{enumerate}
With this definition, $v: {\cal F}(U) \rightarrow \mbox{Hom}(U, {\cal F})$
is a well-defined functor.

Next, in order to show that $u: \mbox{Hom}(U, {\cal F}) \rightarrow
{\cal F}(U)$ is an equivalence of categories, we need to show
that $u \circ v \cong \mbox{Id}$ and $v \circ u \cong \mbox{Id}$.
It is easy to check that $u \circ v = \mbox{Id}_{ {\cal F}(U)}$.

To show that $v \circ u \cong \mbox{Id}$, we shall construct
an invertible natural transformation $\lambda: v \circ u
\Rightarrow \mbox{Id}_{ Hom(U, {\cal F}) }$.

We construct $\lambda$ as follows.
Let $f: U \rightarrow {\cal F}$ be a groupoid morphism
(i.e., an object of $\mbox{Hom}(U, {\cal F})$).
Then $\lambda(f): (v \circ u)(f) \rightarrow f$
is the morphism in $\mbox{Hom}(U, {\cal F})$
(i.e., the natural transformation between 
groupoid morphisms $(v \circ u)(f)$, $f$)
determined as follows.
For any inclusion $\rho: V \hookrightarrow U$ of open $V$ into $U$, define
\begin{displaymath}
\lambda(f)(V): \: (v \circ u)(f) = f(U)|_V \: \longrightarrow
\: f(V)
\end{displaymath}
to be the unique morphism in ${\cal F}$ that makes the
following diagram commute:
\begin{displaymath}
\begin{array}{ccc}
f(U)|_V & \stackrel{ \lambda(f)(V) }{ \longrightarrow } & f(V) \\
\downarrow & & \downarrow
\makebox[0pt][l]{ $\scriptstyle{ f(\rho) }$ } \\
f(U) & = & f(U)
\end{array}
\end{displaymath}
(Existence and uniqueness follow from the second groupoid axiom.)
Since $p_{{\cal F}}( \lambda(f)(V) ) = \mbox{Id}_V$,
the morphism $\lambda(f)(V)$ is invertible.
It is straightforward to check that the $\lambda(f)(V)$
define a natural transformation $\lambda(f): (v \circ u)(f)
\Rightarrow f$,
and it is also straightforward that these
natural transformations $\lambda(f)$ themselves
define an invertible natural transformation $\lambda:
v \circ u \Rightarrow \mbox{Id}$.

Thus, we have that $u \circ v \cong \mbox{Id}$ and
$v \circ u \cong \mbox{Id}$, so $u$ and $v$ are equivalences
of categories.

As a result of this lemma, we now see that giving an object
of ${\cal F}$ over open $U \subseteq X$ is equivalent to specifying
a groupoid morphism $U \rightarrow {\cal F}$.

\subsection{Fiber products of groupoids}

Suppose one has groupoids ${\cal F}_1$, ${\cal F}_2$ and ${\cal G}$
over a space $X$,
together with groupoid morphisms
\begin{eqnarray*}
f_1: & {\cal F}_1 \: \longrightarrow {\cal G} \\
f_2: & {\cal F}_2 \: \longrightarrow {\cal G}
\end{eqnarray*}
Then we can define the fiber product ${\cal F}_1 \times_{{\cal G}}
{\cal F}_2$, as follows.

First, we shall define
the category describing ${\cal F}_1 \times_{{\cal G}} {\cal F}_2$.
Objects of this category are triples $(A_1, A_2, \alpha)$,
where $A_i \in \mbox{Ob } {\cal F}_i$, $p_{{\cal F}_1}(A_1) = 
p_{{\cal F}_2}(A_2)$, and $\alpha: f_1(A_1) \stackrel{\sim}{\longrightarrow}
f_2(A_2)$ an isomorphism in ${\cal G}$.
Morphisms 
\begin{displaymath}
\left( A_1, A_2, \alpha \right) \: \longrightarrow \:
\left( B_1, B_2, \beta \right)
\end{displaymath}
are defined by pairs $(\phi_1, \phi_2)$, where
$\phi_i: A_i \rightarrow B_i$ is a morphism in ${\cal F}_i$,
such that $p_{{\cal F}_1}(\phi_1) = p_{{\cal F}_2}(\phi_2)$,
and $\beta \circ f_1(\phi_1) = f_2(\phi_2) \circ \alpha$.

The projection map $p: {\cal F}_1 \times_{{\cal G}} {\cal F}_2
\rightarrow {\cal U}$ is defined as follows.
\begin{enumerate}
\item On objects,
\begin{displaymath}
p\left( (A_1, A_2, \alpha) \right) \: \equiv \: p_{{\cal F}_1}(A_1)
\: = \: p_{{\cal F}_2}(A_2) 
\end{displaymath}
\item On morphisms,
\begin{displaymath}
p\left( (\phi_1, \phi_2) \right) \: \equiv \: p_{{\cal F}_1}(\phi_1)
\: = \: p_{{\cal F}_2}(\phi_2)
\end{displaymath}
\end{enumerate}
These definitions yield a well-defined functor
$p: {\cal F}_1 \times_{{\cal G}} {\cal F}_2
\rightarrow {\cal U}$.

It is straightforward to check that,
with the definitions above, ${\cal F}_1 \times_{{\cal G}}
{\cal F}_2$ is a well-defined groupoid over $X$.

Now, fibered products are assumed to possess a universal property:
namely, if ${\cal H}$ is any other groupoid over $X$, together
with maps $h: {\cal H} \rightarrow {\cal F}_1$, $k: {\cal H}
\rightarrow {\cal F}_2$ such that the diagram 
\begin{displaymath}
\begin{array}{ccc}
{\cal H} & \stackrel{k}{\longrightarrow} & {\cal F}_2 \\
\makebox[0pt][r]{ $\scriptstyle{h}$ } \downarrow & &
\downarrow \makebox[0pt][l]{ $\scriptstyle{f_2}$ } \\
{\cal F}_1 & \stackrel{f_1}{\longrightarrow} & {\cal G}
\end{array}
\end{displaymath}
commutes, then there exists a unique groupoid morphism
$r: {\cal H} \rightarrow {\cal F}_1 \times_{{\cal G}} {\cal F}_2$
such that $h = \pi_1 \circ r$ and $k = \pi_2 \circ r$.

It is straightforward to check that our fibered products do indeed
satisfy this universal property.  If ${\cal H}$, $h$, and $k$
are above, then define a functor $r: {\cal H} \rightarrow
{\cal F}_1 \times_{{\cal G}} {\cal F}_2$ as follows:
\begin{enumerate}
\item Objects:  Let $A \in \mbox{Ob }{\cal H}$.  Define
$r(A) = \left( h(A), k(A), \mbox{id} \right)$.
\item Morphisms:  Let $\lambda: A \rightarrow B$ be a morphism in
${\cal H}$.  Define $r(\lambda) = \left( h(\lambda), k(\lambda) \right)$.
\end{enumerate}
It is straightforward to check that $r$ possesses the desired
properties, and so our fibered products possess the usual universal
property.

\section{Presheaves on groupoids}     \label{presf}

A presheaf on a groupoid ${\cal F}$ is simply a contravariant functor
from ${\cal F}$.  Note that in the special case the groupoid
${\cal F}$ is a space (i.e., the category is the category of open
sets of some topological space), then this notion of presheaf
coincides with the usual notion of presheaf on a space.


Clearly, the intuition here is that the objects of ${\cal F}$
are open sets, on some sort of generalized space.
If the reader thinks deeply about this intuition, they may
become somewhat confused about how to define a sheaf,
given a presheaf.  After all, to define a sheaf, we need to be
able to talk about coverings of open sets.
Here, because our category ${\cal F}$ of ``open sets'' can
have more interesting morphisms than in the category of open sets
on a standard topological space, it is not quite clear what
a covering of an object of ${\cal F}$ should be.

In the next section we shall speak to this issue.
In order to make sense out of the notion of a sheaf,
we have to specify some additional information, which somehow
captures a notion of ``topology'' of ${\cal F}$. 
This extra information will be precisely a set of coverings of the
objects of ${\cal F}$.

\section{Groupoids and sites}     \label{site}

Just as the total space of a bundle can itself be understood
as a topological space, a groupoid can be understood as a 
``generalized'' topological space.  The notion of generalized
spaces was introduced by the Grothendieck school \cite{artingt,milne},
and basically involves replacing the sets in
point-set topology with categories.  In particular, a generalized
space does not have a {\it set} of points, but rather has a category
of points.  The category of open sets on a generalized space
can have more morphisms than just inclusion maps -- for example,
the open sets can have automorphisms beyond just the identity.

A detailed introduction to the ideas of generalized spaces
is beyond the scope of this paper; see instead
\cite{artingt,milne,mumford1,johnstone,maclane}.
Generalized spaces are typically defined by the category of
maps into them (in algebraic geometry this is referred to as working
with Grothendieck's functor of points).  However, although that approach
is quite powerful, it is rather abstract, and considerably more
general than we shall need.  Thus, in this paper we shall take
a slightly different approach.

In practice, one is often only interested in the
open sets on a generalized space.  The open sets are defined by
some category, which one often wants to possess fibered products.
Now, one would like to define, for example, sheaves on generalized
spaces, for which one needs some notion of a covering of an
open set on a generalized space.
Since there can be more morphisms between the open sets than just
the identity and inclusion maps, and because these morphisms need not
be, in any sense, one-to-one in general,
to specify a topology on a generalized space one must also specify
a set of coverings of each open set, i.e., for each open set
$\eta \in \mbox{Ob } {\cal F}$, one must specify sets
of other objects and morphisms 
\begin{displaymath}
\left\{ \, f_{\alpha}: \: \eta_{\alpha} \: \longrightarrow \: \eta \,
\right\}
\end{displaymath}
where all $\eta_{\alpha} \in \mbox{Ob } {\cal F}$.
Again, if our ``open sets'' were open sets in the usual sense,
then the notion of covering would be clear.  However, because
on a generalized space, the category of open sets can have far
more general morphisms, one must carefully define what is meant by
a covering.  Put another way, because the categorical structure
is more complicated, we must specify coverings in order to capture
any notion of topology.  These coverings are subject to certain consistency
conditions, which we shall describe momentarily.

Given knowledge of the open sets and coverings of open sets
on a generalized space, one can then, for example, define a sheaf
on the generalized space. 

We can also think about these matters in an alternative fashion.
Given some arbitrary category, we can put a topology on the
category by specifying a set of coverings for each object.
The objects of the category are thought of as open sets on
some generalized space. 
In particular, in this fashion we can
think of a groupoid (together with a set of coverings) as defining
a generalized space -- the objects of the groupoid are associated
with the open sets.

A category together with an appropriate set of coverings of the objects
is often known as a {\it site}\footnote{Technically, algebraic
geometers typically use the term site to refer to a category together
with not only a set of coverings, but a sheaf of rings.  As we are
not interested in algebraic structures, we are omitting the customary
sheaves of rings, but have adopted the nomenclature.  We apologize to
any readers offended by our notation.}, i.e., a site is a category with
a topology.  Clearly, as a groupoid is itself described by a category,
we can put a topology on the groupoid (i.e., define a set of coverings),
and make it a site.  Not all sites come from groupoids, however.

A few comments on notation are now in order.
The term site refers to any category with a set of coverings (i.e.,
a topology), and indeed occasionally sites are themselves thought
of as describing categories of open sets on some generalized space
\cite{johnstone,maclane}.  In practice, however, the categorical
structure of a site can be extremely complicated, and so one typically
only refers to those sites coming from groupoids as generalized spaces.
 
In particular, not all sites have an easy understanding
in terms of traditional notions of spaces.  For example, in algebraic
geometry one common site is the site of schemes relative to some
scheme $S$, denoted $Sch/S$.  The objects of this site 
are schemes (together with maps
into the scheme $S$), and the set of coverings is determined by the
appropriate context.  This particular site is quite common
in moduli problems in algebraic geometry, 
whose solutions are often stated in terms 
of sheaves on this site.

Now that we have given an introductory picture of generalized spaces
and sites (i.e., topological categories), we shall discuss more
technical definitions.  Let ${\cal F}$ be some category, which we
wish to interpret as the category of open sets on a generalized space.
(${\cal F}$ might be a groupoid fibered over some other space, for
example.)

Using the definition in \cite{artingt,mumford1}, the category ${\cal F}$
of open sets and the coverings are required to obey the following axioms:
\begin{enumerate}
\item Fibered products $\eta_1 \times_{\xi} \eta_2$ of objects
of ${\cal F}$ exist.
\item $\{ f: \eta' \rightarrow \eta \}$ is a covering if $f$ is 
an isomorphism.  If $\{ f_{\alpha}: \eta_{\alpha} \rightarrow \eta\}$
is a covering and if, for all $\alpha$,
\begin{displaymath}
\left\{ \, f_{\alpha \beta}: \: \eta_{\alpha, \beta} \: \longrightarrow \:
\eta_{\alpha} \, \right\}
\end{displaymath}
is a covering, then the whole collection
\begin{displaymath}
\left\{ \, f_{\alpha} \circ f_{\alpha \beta}: \: \eta_{\alpha, \beta} \:
\longrightarrow \: \eta \, \right\}
\end{displaymath}
is a covering.
\item If $\{ f_{\alpha}: \eta_{\alpha} \rightarrow \eta \}$ is a covering
and $g: \xi \rightarrow \eta$ is any morphism, then
\begin{displaymath}
\left\{ \, \xi \times_{\eta} \eta_{\alpha} \: 
\stackrel{ \pi_1}{\longrightarrow} \:
\xi \, \right\}
\end{displaymath}
is a covering, where $\pi_1$ denotes the projection onto the first factor.
\end{enumerate}

Next, we shall specialize to the case that ${\cal F}$ is a groupoid,
and we shall show how one can naturally put a topology on a groupoid.
First, we shall check that groupoids admit fibered products.
Then, we shall point out a natural notion of covering on a groupoid
over $X$, and check that this notion of covering satisfies the remaining
axioms.

\subsection{Groupoids admit fibered products}

Let $x, y, z \in \mbox{Ob } {\cal F}$, where ${\cal F}$ is a groupoid,
and let $q_x: x \rightarrow z$, $q_y: y \rightarrow z$ be morphisms in 
${\cal F}$.  We shall first show that there exists an object,
which we shall denote $x \times_z y$, together with morphisms
\begin{eqnarray*}
\pi_x: & x \times_z y & \longrightarrow \: x \\
\pi_y: & x \times_z y & \longrightarrow \: y
\end{eqnarray*}
such that the following diagram commutes:
\begin{displaymath}
\begin{array}{ccc}
x \times_z y & \stackrel{ \pi_y }{\longrightarrow} & y \\
\makebox[0pt][r]{ $\scriptstyle{ \pi_x }$ } \downarrow & &
\downarrow \makebox[0pt][l]{ $\scriptstyle{ q_y }$ } \\
x & \stackrel{ q_x }{\longrightarrow} & z
\end{array}
\end{displaymath}

Define $U_x = p_{{\cal F}}(x)$, $U_y = p_{{\cal F}}(y)$, and 
$V = U_x \cap U_y$.  Define $x \times_z y \equiv x|_V$,
and let $\pi_x: x \times_z y \rightarrow x$ be the canonical inclusion map.
Let $\pi_y: x \times_z y \rightarrow y$ be the unique morphism
such that $q_y \circ \pi_y = q_x \circ pi_x$ (whose existence and
uniqueness are guaranteed by the second groupoid axiom).
We now have the desired object $x \times_z y$. 

We still need to check that the object $x \times_z y$ has the
desired universality property.  Let $w \in \mbox{Ob } {\cal F}$,
together with morphisms $u: w \rightarrow x$, $v: w \rightarrow y$
such that the diagram 
\begin{displaymath}
\begin{array}{ccc}
w & \stackrel{v}{\longrightarrow} & y \\
\makebox[0pt][r]{ $\scriptstyle{ u }$ } \downarrow & &
\downarrow \makebox[0pt][l]{ $\scriptstyle{ q_y }$ } \\
x & \stackrel{ q_x }{\longrightarrow} & z
\end{array}
\end{displaymath}
commutes.
We shall now demonstrate the existence of a unique morphism
$t: w \rightarrow x \times_z y$ such that $\pi_y \circ t = v$ 
and $\pi_x \circ t = u$.

Define $U_w = p_{{\cal F}}(w)$.  Since there are morphisms
$w \rightarrow x$ and $w \rightarrow y$, we know that
$U_w \subseteq U_x \cap U_y = V$.  Define $h: U_w \hookrightarrow V$.

From the second groupoid axoim, there exists a unique morphism
$t_1: w \rightarrow x \times_z y$ such that $p_{{\cal F}}(t_1) = h$,
and $\pi_x \circ t_1 = u$.
Similarly, there exists a unique morphism $t_2: w \rightarrow x \times_z y$
such that $p_{{\cal F}}(t_2) = h$ and $\pi_y \circ t_2 = v$.

Finally, again from the second groupoid axiom, there exists a unique
morphism $t: w \rightarrow x \times_z y$ such that $p_{{\cal F}}(t) = h$
and $q_x \circ \pi_x \circ t = q_x \circ u$.
But we also know that $t_1$ satisfies the same hypothesis,
hence by uniqueness, $t = t_1$.  Using the facts that
$q_x \circ \pi_x = q_y \circ \pi_y$ and $q_x \circ u = q_y \circ v$,
one also finds from uniqueness that $t = t_2$.

Hence, we have a morphism $t: w \rightarrow x \times_z y$ with the
desired commutivity properties, and so our fibered product
$x \times_z y$ does indeed possess the necessary universality property.
Thus, groupoids admit fibered products.

\subsection{Coverings for groupoids}

A natural notion of covering for a groupoid ${\cal F}$ is as follows.
Define a collection of maps $\{ f_{\alpha}: \eta_{\alpha} \rightarrow \eta\}$
to be a covering of $\eta \in \mbox{Ob } {\cal F}$,
precisely when the collection of open sets $\{ p_{{\cal F}}(\eta_{\alpha}) \}$
defines a covering (in $X$) of the open set $p_{{\cal F}}(\eta)
\subseteq X$.

Then, the axioms for a set of coverings to define a topology
are trivial to check.

First, $\{ g: \eta' \rightarrow \eta \}$ will define a covering
precisely when $p_{{\cal F}}(\eta') = p_{{\cal F}}(\eta)$,
i.e., precisely when $g$ is an isomorphism.

Next, suppose $\{ f_{\alpha}: \eta_{\alpha} \rightarrow \eta \}$
is a covering and, for all $\alpha$, $\{ f_{\alpha \beta}:
\eta_{\alpha, \beta} \rightarrow \eta_{\alpha} \}$ is a covering also.
Then certainly $\{ p_{{\cal F}}(\eta_{\alpha, \beta} ) \}$
is an open cover of $p_{{\cal F}}(\eta) \subseteq X$,
so the whole collection
\begin{displaymath}
\left\{ \, f_{\alpha} \circ f_{\alpha \beta}: \:
\eta_{\alpha, \beta} \: \longrightarrow \: \eta \, \right\}
\end{displaymath}
is a covering.

Finally, suppose that $\{ f_{\alpha}: \eta_{\alpha} \rightarrow \eta \}$
is a covering, and $g: \xi \rightarrow \eta$ is any morphism.
Then $p_{{\cal F}}(\xi) \subseteq p_{{\cal F}}(\eta)$,
so $\{ p_{{\cal F}}(\xi) \cap p_{{\cal F}}(\eta_{\alpha}) \}$ 
is an open cover of $p_{{\cal F}}(\xi)$,
hence 
\begin{displaymath}
\left\{ \, \xi \times_{\eta} \eta_{\alpha} \: \stackrel{\pi_1}{
\longrightarrow} \: \xi \, \right\}
\end{displaymath}
is a covering.

\section{Stacks}    \label{stk}

Now that we have discussed groupoids, we shall discuss
under what circumstances groupoids are stacks.

As noted earlier, a groupoid is equivalent to a prestack.
So, a stack is merely a special groupoid, one satisfying certain
gluing axioms.

In particular, a stack is a groupoid that satisifes two gluing
axioms:
\begin{enumerate}
\item Gluing for morphisms:  The presheaf $\underline{\mbox{Hom}}_U(x,y)$
of sets must be a sheaf of sets, for every open $U \subseteq X$ and
every $x, y \in \mbox{Ob } {\cal F}(U)$.
\item Gluing for objects: 
Let $U$ be any open set in $X$, and $\{ U_{\alpha} \}$ any
open cover of $U$.  Let $\{ x_{\alpha} \in \mbox{Ob } {\cal F}(U_{\alpha}) \}$
be a family of objects over the elements of $\{ U_{\alpha} \}$,
and let $\phi_{\alpha \beta}: x_{\beta}|_{U_{\alpha \beta}} \stackrel{\sim}{
\longrightarrow} x_{\alpha}|_{U_{\alpha \beta}}$ be a family of isomorphisms
between the restrictions of the $\{ x_{\alpha} \}$.
Assume that $\phi_{\alpha \alpha} = \mbox{Id}$, and that the following
diagram commutes:
\begin{displaymath}
\begin{array}{ccccc}
x_{\gamma} |_{U_{\beta \gamma}} & \stackrel{\phi_{\beta \gamma}}{
\longrightarrow} & x_{\beta}|_{U_{\beta \gamma}} & \longrightarrow &
x_{\beta} \\
\downarrow & & & & \uparrow \\
x_{\gamma} & & & & x_{\beta}|_{U_{\alpha \beta}} \\
\uparrow & & & & \downarrow \makebox[0pt][l]{ $\scriptstyle{ \phi_{\alpha
\beta} }$ } \\
x_{\gamma}|_{U_{\alpha \gamma}} & & & & x_{\alpha} |_{U_{\alpha \beta}} \\
\makebox[0pt][r]{ $\scriptstyle{ \phi_{\alpha \gamma} }$ } \downarrow &
& & & \downarrow \\
x_{\alpha} |_{U_{\alpha \gamma}} & & \longrightarrow & & x_{\alpha} 
\end{array}
\end{displaymath}
where unmarked arrows are canonical.
(Briefly, ``$\phi_{\alpha \beta} \circ \phi_{\beta \gamma} = 
\phi_{\alpha \gamma}$.'')
There there must exist $x \in \mbox{Ob } {\cal F}(U)$
together with isomorphisms $\psi_{\alpha}: x|_{\alpha} \rightarrow x_{\alpha}$
such that the following diagram commutes:
\begin{displaymath}
\begin{array}{ccccc}
x|_{U_{\beta}} & \longrightarrow & x & \longleftarrow & x|_{U_{\alpha}} \\
\makebox[0pt][r]{ $\scriptstyle{ \psi_{\beta} }$ } \downarrow & & & &
\downarrow \makebox[0pt][l]{ $\scriptstyle{ \psi_{\alpha} }$ } \\
x_{\beta} & & & & x_{\alpha} \\
\uparrow & & & & \uparrow \\
x_{\beta}|_{U_{\alpha \beta}} & & \stackrel{ \phi_{\alpha \beta} }{
\longrightarrow} & & x_{\alpha} |_{U_{\alpha \beta}} 
\end{array}
\end{displaymath}
where unmarked arrows are canonical.
(Briefly, ``$\phi_{\alpha \beta} \circ \psi_{\beta} = \psi_{\alpha}$.'')
\end{enumerate}

It is straightforward to check that if a groupoid satisfies these 
axioms, then an associated prestack satisfies the axioms of
\cite{dt2} to be a stack.  Also, conversely, given a prestack that
happens to be a stack, the associated groupoid satisfies the axioms to
be a stack.

\section{Sheaves on stacks}    \label{sfstk}

\subsection{Presheaves on groupoids}

Before studying sheaves on stacks, we shall warm up by
considering presheaves on groupoids.  Just as a presheaf of, say,
sets, on a space $X$ is a contravariant functor from the category
of open sets to $\underline{\mbox{Set}}$, a presheaf of sets on
a groupoid ${\cal F}$ over $X$ \cite{mumford1} is a contravariant functor from
${\cal F}$ to $\underline{\mbox{Set}}$.

\subsection{Sheaves on topological groupoids}

In addition to presheaves, we can also define sheaves on groupoids.
Now, a sheaf differs from a presheaf through possessing a glueing
property, which is phrased in terms of open covers.  So, in order to
talk about sheaves on groupoids, one must choose a set of coverings
 -- one must put a topology on the groupoid.

Let us assume a topology has been chosen on a groupoid ${\cal F}$ over
$X$.  Let ${\cal S}$ be a presheaf on ${\cal F}$.
Then, ${\cal S}$ is a sheaf, not just a presheaf, if \cite{mumford1} for 
all objects $\eta \in \mbox{Ob } {\cal F}$, and for all
covers $\{ \eta_{\alpha} \stackrel{\rho_{\alpha}}{\longrightarrow}
\eta \}$ of $\eta$,
if $\{ s_{\alpha} \in {\cal S}(\eta_{\alpha}) \}$ is a family of
elements such that
\begin{displaymath}
\rho^{\alpha *}_{\alpha \beta} s_{\alpha} \: = \:
\rho^{\beta *}_{\alpha \beta} s_{\beta}
\end{displaymath}
(where $\rho^{\alpha}_{\alpha \beta}$, $\rho^{\beta}_{\alpha \beta}$
are the projection maps $\eta_{\alpha} \times_{\eta} \eta_{\beta}
\rightarrow \eta_{\alpha}, \eta_{\beta}$),
then there exists a unique $s \in {\cal S}(\eta)$ such that
$\rho_{\alpha}^* s = s_{\alpha}$.

More briefly, the gluing condition for a presheaf to be a sheaf
over a groupoid is precisely analogous to the usual gluing
condition.

\subsection{D-brane ``bundles''}

As is well-known, on the worldvolume of $N$ coincident D-branes,
there is a $U(N)$ bundle.  However, in the presence of a nontrivial
$B$ field, this story is slightly modified.  Specifically, if we
describe a nontrivial $B$ field in terms of local coordinate
data $(B^{\alpha}, A^{\alpha \beta}, h_{\alpha \beta \gamma})$
(as described in, for example, \cite{hitchin}) then the transition
functions $g_{\alpha \beta}$ for the ``bundle'' obey
\begin{displaymath}
g_{\alpha \beta} g_{\beta \gamma} g_{\gamma \alpha} \: = \:
h_{\alpha \beta \gamma} I
\end{displaymath}
(where $I$ is the $N \times N$ identity matrix) on triple overlaps
\cite{freeded}.

Such twisted bundles can be described in terms of certain sheaves
on stacks.  In particular, just as gauge fields are naturally understood
in terms of connections on bundles, $B$ fields are naturally understood
in terms of connections on gerbes, which are special kinds of stacks.
This is described in more detail in, for example, \cite{brylinski,dt2};
we shall not repeat the story here, but shall instead assume
the reader is acquainted with it.

Before describing how D-brane ``bundles'' are certain sheaves
on stacks, we shall first briefly review how one finds $h_{\alpha
\beta \gamma}$ in terms of stacks.  Let ${\cal F}$ be a stack
which happens to be a gerbe, and let $\{ U_{\alpha} \}$ be a
good open cover over $X$, and let $\{ \eta_{\alpha} \in
\mbox{Ob } {\cal F}(U_{\alpha}) \}$ be a collection of objects lying
over elements of the open cover.  Let $u_{\alpha \beta}:
\eta_{\alpha} |_{U_{\alpha \beta}} \stackrel{\sim}{\longrightarrow}
\eta_{\beta} |_{U_{\alpha \beta}}$ be a collection of isomorphisms.
(For a stack which is a gerbe, all objects lying over a contractible
open set are isomorphic.)  Let $\overline{u}_{\alpha \beta}:
\eta_{\alpha} |_{U_{\alpha \beta \gamma}} \stackrel{\sim}{\longrightarrow}
\eta_{\beta} |_{U_{\alpha \beta \gamma}}$ be associated isomorphisms,
as described in \cite{dt2}.
Then, 
\begin{displaymath}
h_{\alpha \beta \gamma} \: = \:
\overline{u}_{\gamma \alpha} \circ \overline{u}_{\beta \gamma} \circ
\overline{u}_{\alpha \beta} \: \in \:
\mbox{Aut}\left( \eta_{\alpha} |_{U_{\alpha \beta \gamma}} \right)
\end{displaymath}
(For a stack which is a gerbe, and in particular a gerbe with band
$C^{\infty}(U(1))$, automorphisms of objects over $U$ are in 1-1
correspondence with elements of $C^{\infty}(U, U(1))$.)

Now, consider sheaves ${\cal S}$ on ${\cal F}$ of the following form.
First, for all open $U \subseteq X$,
for all objects $\eta \in \mbox{Ob } {\cal F}(U)$,
let ${\cal S}(\eta)$ be either empty or $C^{\infty}(U, U(N))$.
Second, if $\varphi$ is an automorphism of $\eta$, isomorphic
to some element of $C^{\infty}(U, U(1))$ which we shall also denote by
$\varphi$, then ${\cal S}(\varphi)$ acts on ${\cal S}(\eta)$ by
multiplying by $\varphi \times \varphi \times \cdots \times \varphi
 = \varphi I$, where $I$ is the $N \times N$ identity matrix.

Next, we shall check that for a sheaf on ${\cal F}$ of this form,
transition functions are twisted on triple overlaps just as for
``bundles'' on D-branes.

Let $\{ s_{\alpha} \in {\cal S}(\eta_{\alpha}) \}$ be a collection
of local sections.
Define
\begin{eqnarray*}
g_{\alpha \beta} & = & \left[ s_{\beta} |_{U_{\alpha \beta \gamma}} \right]
\, \left[ {\cal S}( \overline{u}_{\alpha \beta} )( s_{\alpha}|_{U_{
\alpha \beta \gamma}} ) \right]^{-1} \\
 & \in & {\cal S}\left( \eta_{\beta}|_{U_{\alpha \beta}} \right)
\end{eqnarray*}
On triple overlaps, it is straightforward to calculate that
\begin{displaymath}
\left[ g_{\gamma \alpha} \right] \,
\left[ {\cal S}( \overline{u}_{\gamma \alpha} )( g_{\beta \gamma} ) \right] \,
\left[ {\cal S}( \overline{u}_{\gamma \alpha} \circ \overline{u}_{
\beta \gamma} )( g_{\alpha \beta} ) \right]
\: = \: h_{\alpha \beta \gamma} I
\end{displaymath}
where $I$ is the $N \times N$ identity matrix.

\section{Conclusions}     \label{concl}

In this paper, we have explicitly observed that the twisted ``bundles''
on D-branes can be understood in terms of sheaves on stacks.
We have also taken this opportunity to provide a readable overview
of stacks and generalized spaces, in order to aid the reader
in grappling with the notion of a sheaf on a stack.

Since the twisted ``bundle'' can be equivalently understood
as an ordinary bundle on a stack, one is led to ask some interesting
questions about string compactifications and these generalized spaces.
For example, to what extent can one compactify strings on generalized
spaces?  Can ``turning on the B field'' be equivalently understood
as compactifying on a generalized space?
We hope to answer these questions in future work.

\section{Acknowledgements}

We would like to thank D. Ben-Zvi for originally insisting to us
that it should be possible to understand the twisted ``bundles''
on D-branes in terms of sheaves on stacks, and for useful comments
on a rough draft.  We would also like to
thank P.~Aspinwall, A.~Knutson, D.~Morrison, and R.~Plesser for useful
conversations.

\appendix

\section{Equivalence of groupoids and prestacks}    \label{gpdeqprest}

In this appendix we shall show that the notion of
groupoid is equivalent to the notion of prestack
as defined in \cite{dt2} (with the extra hypothesis on
prestacks that all morphisms are invertible).

\subsection{Prestacks are groupoids}

Next,
we shall show that this notion of a groupoid over a space
is almost identical to the notion of a prestack on a space
as defined in \cite{dt2}.  

Let ${\cal C}$ be a prestack (or presheaf of categories)
on a space $X$, as defined in
\cite{dt2}.  We shall also assume that for every open set $U \subseteq X$,
all of the morphisms in the category ${\cal C}(U)$ are invertible.
(This is one of the axioms appearing in the definition of gerbe
in \cite{dt2}.)  This additional axiom will only be needed to
prove the second groupoid axiom.

In order to derive a groupoid from this prestack ${\cal C}$,
we first need to find a category ${\cal F}$ and a functor $p_{{\cal F}}: 
{\cal F} \rightarrow
{\cal C}$, and then demonstrate that ${\cal F}$ and $p_{{\cal F}}$ satisfy the
appropriate axioms.

Define ${\cal F}$ as follows.  
\begin{enumerate}
\item Objects of ${\cal F}$ are pairs $(U, \xi)$
where $\xi \in \mbox{Ob } {\cal C}(U)$, $U \in \mbox{Ob } {\cal U}$.
\item A morphism $(U, \xi) \rightarrow (V, \zeta)$ is a pair of
maps $(\rho, f)$, where $\rho: U \hookrightarrow V$
and $f: \xi \rightarrow \rho^* \zeta$ is a morphism in ${\cal C}(U)$.
\end{enumerate}
Given two morphisms 
\begin{eqnarray*}
(\rho_1, f_1): \: (U_2, \xi_2) \: \longrightarrow \: (U_1, \xi_1) \\
(\rho_2, f_2): \: (U_3, \xi_3) \: \longrightarrow \: (U_2, \xi_2)
\end{eqnarray*}
we define their composition to be
\begin{displaymath}
(\rho_1, f_1) \circ (\rho_2, f_2) \: = \:
( \rho_1 \rho_2, \varphi_{1,2}^{-1} \circ \rho_2^* f_1 \circ f_2 )
\end{displaymath}
where $\varphi$ is the invertible natural transformation appearing
in the definition of prestack.
It can be shown that this composition is associative.

Define the functor $p_{{\cal F}}: {\cal F} \rightarrow {\cal U}$ as follows.
\begin{enumerate}
\item On objects, $p_{{\cal F}}( (U, \xi) ) = U$
\item On morphisms, $p_{{\cal F}}( (\rho, f) ) = \rho$
\end{enumerate}

It can be shown that with these definitions of ${\cal F}$ and $p_{{\cal F}}$,
the two axioms for a groupoid are satisfied.  
(In order to show the second axiom for a groupoid, we must use
the assumption that all morphisms in ${\cal C}$ are invertible.)

\subsection{Groupoids are prestacks}

Let ${\cal F}$ be a groupoid over a space $X$.
We shall now show how ${\cal F}$ defines a prestack ${\cal C}$
on $X$.

First, for any open set $U \subseteq X$, define ${\cal C}(U) = {\cal F}(U)$.
(Note that this means that all morphisms in ${\cal C}$ are invertible,
a slightly stronger condition than needed for a prestack.)

Next, for each inclusion $\rho: V \hookrightarrow U$ of open sets,
define a functor $\rho^*: {\cal C}(U) \rightarrow {\cal C}(V)$ as follows.
\begin{enumerate}
\item Objects:  Let $x \in \mbox{Ob } {\cal C}(U)$.
Define $\rho^* x$ to be $x |_V$.
\item Morphisms:  Let $x, y \in \mbox{Ob } {\cal C}(U)$,
and $f: x \rightarrow y$ a morphism in ${\cal C}(U)$.
Let $g_x: x|_V \rightarrow x$, $g_y: y|_V \rightarrow y$ denote the
canonical maps.  Then by the second groupoid axiom,
there exists a unique morphism, call it $\rho^* f$,
mapping $x|_V \rightarrow y|_V$, such that the following diagram commutes:
\begin{displaymath}
\begin{array}{ccc}
x & \stackrel{f}{\longrightarrow} & y \\
\makebox[0pt][r]{ $\scriptstyle{ g_x }$ } \uparrow & &
\uparrow \makebox[0pt][l]{ $\scriptstyle{ g_y }$ } \\
x|_V & \stackrel{ \rho^* f }{\longrightarrow } & y|_V
\end{array}
\end{displaymath}
\end{enumerate}
It is straightforward to check that with these definitions,
$\rho^*: {\cal C}(U) \rightarrow {\cal C}(V)$ is a well-defined functor.

Finally, for each pair $\rho_1: U_2 \hookrightarrow U_1$,
$\rho_2: U_3 \hookrightarrow U_2$ of composable inclusions of open sets,
we define an invertible natural transformation
\begin{displaymath}
\varphi_{\rho_1, \rho_2}: \: (\rho_1 \rho_2)^* \: \Longrightarrow \:
\rho_2^* \circ \rho_1^*
\end{displaymath}
as follows.
Let $x \in \mbox{Ob } {\cal C}(U_1)$, and let
\begin{eqnarray*}
g_2: & x|_{U_2} & \longrightarrow \: x \\
g_3: & x|_{U_3} & \longrightarrow \: x \\
g_{23}: & x|_{U_2}|_{U_3} & \longrightarrow \: x|_{U_2}
\end{eqnarray*}
be the canonical maps.
Then, define $\varphi(x): x|_{U_3} \rightarrow x|_{U_2}|_{U_3}$
is defined to be the unique morphism that makes the following
diagram commute:
\begin{displaymath}
\begin{array}{ccc}
x|_{U_3} & \stackrel{ g_3 }{\longrightarrow} & x \\
\makebox[0pt][r]{ $\scriptstyle{ \varphi(x) }$ } \downarrow & &
\uparrow \makebox[0pt][l]{ $\scriptstyle{ g_2 }$ } \\
x|_{U_2}|_{U_3} & \stackrel{ g_{23} }{\longrightarrow } & x|_{U_2}
\end{array}
\end{displaymath}
(whose existence is guaranteed by the second groupoid axiom).
It is straightforward to check that with the definition above,
$\varphi$ is a natural transformation, and moreover satisfies the
condition that for any three composable inclusions, the following
diagram commutes:
\begin{displaymath}
\begin{array}{ccc}
(\rho_1 \rho_2 \rho_3)^* & \Longrightarrow & \rho_3^* \circ 
( \rho_1 \rho_2)^* \\
\Downarrow & & \Downarrow \\
(\rho_2 \rho_3)^* \circ \rho_1^* & \Longrightarrow &
\rho_3^* \circ \rho_2^* \circ \rho_1^*
\end{array}
\end{displaymath}

With the definitions above, the set of data $( {\cal C}(U),
\rho^*, \varphi)$ defines a prestack.

The attentive reader will correctly note that we have not uniquely
defined a prestack from a groupoid, as the object $x|_U$ is only
defined up to unique isomorphism.  However, it is straightforward
to check that any two prestacks obtained from the same groupoid,
and differing only in the precise choices of $x|_U$, are isomorphic.
Specifically, let ${\cal C}_1$ and ${\cal C}_2$ denote two 
prestacks obtained from the same groupoid ${\cal F}$ on $X$,
and differing as above.
Define a Cartesian functor $F: {\cal C}_1 \rightarrow {\cal C}_2$
as follows.
For any open $U$, since ${\cal C}_1(U) = {\cal C}_2(U) = {\cal F}(U)$,
define $F(U)$ to be the identity functor.
Then, for any inclusion $\rho: V \hookrightarrow U$,
define a natural transformation $\chi_{\rho}$ as follows.
For any $x \in \mbox{Ob } {\cal C}_1(U)$, define $\chi_{\rho}(x)$
to be the unique map $\rho_{{\cal C}_2}^* (x) \rightarrow \rho_{{\cal C}_1}^*
(x)$ such that the following diagram commutes:
\begin{displaymath}
\begin{array}{ccc}
\rho_{{\cal C}_2}^*(x) & \stackrel{ \chi_{\rho}(x) }{\longrightarrow} &
\rho_{{\cal C}_1}^*(x) \\
\downarrow & & \downarrow \\
x & = & x
\end{array}
\end{displaymath}
where the unmarked arrows are canonical.
(Existence and uniqueness follow from the second groupoid axiom.)
It is straightforward to check that $\chi_{\rho}$ is a natural
transformation.
It is also straightforward to check that for any pair $\rho_2: U_3
\hookrightarrow U_2$, $\rho_1: U_2 \hookrightarrow U_1$
of composable inclusions of open sets, the following diagram commutes:
\begin{displaymath}
\begin{array}{ccccc}
\rho_{2 {\cal C}_2}^* \circ \rho_{1 {\cal C}_2}^* \circ F(U_1)
& \stackrel{ \chi_1 }{\Longrightarrow} &
\rho_{2 {\cal C}_2}^* \circ F(U_2) \circ \rho_{1 {\cal C}_1}^*
& \stackrel{ \chi_2 }{\Longrightarrow} &
F(U_3) \circ \rho_{2 {\cal C}_1}^* \circ \rho_{1 {\cal C}_1}^* \\
\makebox[0pt][r]{ $\scriptstyle{ \varphi^{{\cal C}_2} }$ } \Uparrow
& & & & \Uparrow
\makebox[0pt][l]{ $\scriptstyle{ \varphi^{{\cal C}_1} }$ } \\
( \rho_1 \rho_2)_{{\cal C}_2}^* \circ F(U_1) & &
\stackrel{\chi_{12}}{\Longrightarrow} & &
F(U_3) \circ (\rho_1 \rho_2)_{{\cal C}_1}^*
\end{array}
\end{displaymath}
In other words, the natural transformations $\chi_{\rho}$
are compatible with the functors $F(U)$, and
so $( F, \chi_{\rho})$ defines a Cartesian functor
(in the sense of \cite{dt2}).
Clearly, this Cartesian functor defines an equivalence of 
the prestacks ${\cal C}_1$ and ${\cal C}_2$.

\subsection{Cartesian functors and morphisms of groupoids}

Recall that a morphism of groupoids ${\cal F} \rightarrow {\cal G}$ 
over a space $X$
is defined
to be a functor $F: {\cal F} \rightarrow {\cal G}$ such that
$p_{{\cal G}} \circ F = p_{{\cal F}}$, i.e.,
\begin{displaymath}
\begin{array}{ccc}
{\cal F} & \stackrel{ F }{\longrightarrow} & {\cal G} \\
\makebox[0pt][r]{ $\scriptstyle{ p_{{\cal F}} }$ } \downarrow & &
\downarrow \makebox[0pt][l]{ $\scriptstyle{ p_{{\cal G}} }$ } \\
X & = & X
\end{array}
\end{displaymath}
commutes.

A Cartesian functor between prestacks, as defined in \cite{dt2},
defines a morphism between the associated groupoids, as we shall
now demonstrate.

Let ${\cal C}$, ${\cal D}$ be two prestacks (as defined in \cite{dt2}),
subject to the additional constraint that all morphisms in all categories
${\cal C}(U)$ and ${\cal D}(U)$ are invertible.
Let $(F, \chi): {\cal C} \rightarrow {\cal D}$ be a Cartesian functor.
Define a functor $F: {\cal F} \rightarrow {\cal G}$ as follows:
\begin{enumerate}
\item Objects:  Let $(U, \xi) \in \mbox{Ob } {\cal F}$, i.e.,
$\xi \in \mbox{Ob } {\cal C}(U)$.
Define 
\begin{displaymath}
F( (U, \xi)) \: = \: (U, F(U)(\xi))
\end{displaymath}
\item Morphisms:  Let $(\rho, f): (U, \xi) \rightarrow (V, \zeta)$,
i.e., $f: \xi \rightarrow \zeta|_U$.
Define
\begin{displaymath}
F( (\rho,f) ) \: = \: \left(\rho, \: \chi_{\rho}^{-1} \circ F(U)(f) \right)
\end{displaymath}
\end{enumerate}

It can be shown that the functor $F$ is well-defined, and moreover
$p_{{\cal G}} \circ F = p_{{\cal F}}$.

Conversely, let $f: {\cal F} \rightarrow {\cal G}$ be a morphism of
groupoids over $X$, and let ${\cal C}$, ${\cal D}$ denote
prestacks associated to ${\cal F}$, ${\cal G}$, respectively.
Define a Cartesian functor $(F, \chi_{\rho}): {\cal C} \rightarrow
{\cal D}$ as follows:
\begin{enumerate}
\item For every open set $U \subseteq X$, the functor $F(U):
{\cal C}(U) \rightarrow {\cal D}(U)$ is induced by $f$,
as ${\cal C}(U) = {\cal F}(U)$ and ${\cal D}(U) = {\cal G}(U)$.
\item Let $\rho: V \hookrightarrow U$ be an inclusion of open sets
in $X$.  For any $x \in \mbox{Ob } {\cal F}(U)$, define
$\chi_{\rho}(x)$ to be the unique morphism that makes the following
diagram commute:
\begin{displaymath}
\begin{array}{ccc}
f(x)|_V & \stackrel{ \chi_{\rho}(x) }{\longrightarrow} & f\left( x|_V \right)\\
\downarrow & & \downarrow \\
f(x) & = & f(x) 
\end{array}
\end{displaymath}
where unmarked arrows are (images of) canonical maps.
(Existence and uniqueness follow from the second groupoid axiom.)
It is straightforward to check that this defines a natural
transformation $\chi_{\rho}: \rho_{{\cal D}}^* \circ F(U) \Rightarrow
F(V) \circ \rho_{{\cal C}}^*$, and moreover that this natural
transformation makes $(F, \chi_{\rho})$ into a Cartesian functor.
\end{enumerate}

\subsection{Natural transformations and 2-arrows}

In the previous section, we argued that a Cartesian functor
$F: {\cal C} \rightarrow {\cal D}$ between prestacks
(in the sense of \cite{dt2}) defines a groupoid morphism
between the groupoids ${\cal F}$, ${\cal G}$ associated to
${\cal C}$ and ${\cal D}$, respectively.
In this section we shall point out that a 2-arrow
$\Psi: F \rightarrow G$ (defined as in \cite{dt2})
between Cartesian functors
$F, G: {\cal C} \rightarrow {\cal D}$ defines a natural
transformation $\eta: F \rightarrow G$ between the associated
groupoid morphisms.

Let $(U, \xi) \in \mbox{Ob } {\cal F}$, i.e., $\xi \in
\mbox{Ob } {\cal C}(U)$.  Then, we define the natural
transformation $\eta$ by
\begin{displaymath}
\eta\left( (U, \xi) \right) \: \equiv \:
\mbox{Id}_U \times \Psi(U)(\xi): \:
\left( U, F(U)(\xi) \right) \: \longrightarrow \:
\left( U, G(U)(\xi) \right)
\end{displaymath}

It is straightforward to check that $\eta$ is a well-defined
natural transformation.

Conversely, let $f, g: {\cal F} \rightarrow {\cal G}$ be a pair
of morphisms of groupoids over $X$, and let $\eta: f \Rightarrow g$
be a natural transformation relating them.
Let ${\cal C}$, ${\cal D}$ be prestacks associated to ${\cal F}$,
${\cal G}$, and let $F, G: {\cal C} \rightarrow {\cal D}$ 
be Cartesian functors associated to $f$, $g$, respectively.
We shall now define a 2-arrow $\Psi: F \Rightarrow G$,
in the sense of \cite{dt2}.

For any open set $U$, recall that the functors $F(U)$ and $G(U)$ are
induced directly from $f$ and $g$.  Clearly, $\eta$ induces
a natural transformation $\Psi(U): F(U) \Rightarrow G(U)$.
Also, for any inclusion $\rho: V \hookrightarrow U$ of open sets,
it is straightforward to check that the following diagram commutes:
\begin{displaymath}
\begin{array}{ccc}
\rho_{{\cal D}}^* \circ F(U) & \stackrel{ \Psi(U) }{\Longrightarrow} &
\rho_{{\cal D}}^* \circ G(U) \\
\makebox[0pt][r]{ $\scriptstyle{ \chi^F_{\rho} }$ } \Downarrow & &
\Downarrow \makebox[0pt][l]{ $\scriptstyle{ \chi^G_{\rho} }$ } \\
F(V) \circ \rho_{{\cal C}}^* & \stackrel{ \Psi(V) }{\Longrightarrow} &
G(V) \circ \rho_{{\cal C}}^*
\end{array}
\end{displaymath}
Thus, we have defined a 2-arrow $\Psi: F \Rightarrow G$.

\end{document}